%% file: bulge.tex
\input ./macros

\input ./title

\section 1. Introduction

Kinematic surveys of a population of discrete objects are an
increasingly important kind of data in galactic astronomy. The objects
may be stars in globular clusters or elsewhere in the Galaxy (e.g.,
Meylan \& Mayor 1986), emission line objects in the Galaxy or other
galaxies (e.g., Ciardullo \etal\ 1993, Hui 1993, Arnaboldi \etal\
1994, Tremblay \etal\ 1995, Beaulieu 1996, Sevenster \etal\ 1997a,b), or
galaxies in a cluster (e.g., Colless \& Dunn 1996).  Such surveys
usually measure sky positions and line-of-sight velocities, but for
some systems proper motions are also available (e.g., Spaenhauer
\etal\ 1992).

One would like to be able to throw these data at some dynamical
analysis machine and reap all the dynamical results implicit in the
data, but there is no such machine. Some progress towards this goal
has been made, notably by D.~Merritt and collaborators (see
Merritt 1993, Merritt \& Tremblay 1993, Merritt \& Gebhardt 1994, and
especially Merritt 1996). These papers develop methods for
reconstructing mass profiles (including dark matter) from kinematical
observations, in a model independent way. But at present they extend
only to axisymmetric systems viewed in the equatorial plane.  So for
triaxial systems, and certainly for non-equilibrium systems like
clusters of galaxies, it is basically \Nbody\ simulations that have to
be confronted with data. How can we best do this quantitatively?

Generally speaking, there are three questions one would like
answers to when comparing \Nbody\ models with observations.
\item{(i)} How should a model be scaled and oriented to best fit the
data?
\item{(ii)} Could the data at hand have plausibly come from a particular
model's distribution, or do the data rule out the model?
\item{(iii)} If there are several plausible models, which one do the
data favor?
\par\noindent All three are answerable if we can calculate
the likelihood function, which is the probability of having gathered
the actual data under a particular model. Suppose for definiteness
that the data consist of measurements of sky position $l,b$, and
line-of-sight velocity $v$, with negligible errors. Let us also assume
for now that the simulation is so fine grained that it effectively
gives us a distribution function $f$. One usually thinks of $f$ as a
function of phase space variables, but we can change variables to
express it as $f(l,b,v,\eta)$, where $\eta$ stands for three
unmeasured numbers (e.g., distance and proper motion). Then the
probability of drawing values $l_k,b_k,v_k$ from $f$ is
$$ \cprob(l_k,b_k,v_k|f) = \int\!f(l_k,b_k,v_k,\eta)\,d\eta.
   \putnum$$
Assuming the data on different objects are independent, we have for
the likelihood:
$$ \likeli = \prod_k\left[\int\!f(l_k,b_k,v_k,\eta)\,d\eta\right].
   \putnum$$  \nameeq{likelicont}
Since $f(l,b,v,\eta)$ will depend on the scalings and orientation
adopted, we can fit for these parameters---the peak of $\likeli$ in
the relevant parameter space estimates the parameters and the
broadness of that peak gives uncertainties.  To answer question (ii),
we can test if the value of $\likeli$ is typical of random data sets
drawn for that $f$; if $\likeli$ is anomalously low, we can infer that
$f$ is inconsistent with the data.  Question (iii) can be answered by
comparing $\likeli$ for the various models available; there is an
extra complication though, in that we must marginalize over the
parameters for each model---see Sivia (1996) for a discussion of this
point.  I will not address model comparison in this paper.

The contribution of this paper is to derive and test a practical
approximation to the `in-principle' procedure above. We need an
approximation because particle simulations do not give us $f$
directly; we need to smooth somehow.  Smoothings in general introduce
biases, so we have to monitor for biases and correct for them if
necessary.  But bearing that caution in mind, the smoothing I propose
to use is the simple-minded one of just binning in $l,b,v$, i.e.,
assuming that $f$ is constant within boxes in $l,b,v$ space. Let us
say that for some choice of scaling and orientation parameters, the
$i$-th bin has $m_i$ model points and $s_i$ data object points; also
let $M=\sum_im_i$ and $S=\sum_is_i$.  This immediately suggests
minimizing $\chi^2$ to obtain a best fit, but that is a bad
idea. Minimizing $\chi^2$ implicitly assumes that the $s_i$ follow a
Gaussian distribution, the mean and variance in this case being both
equal to $m_iS/M$. This is fine if all the $s_i\gg1$, but $S$ being
typically dozens to hundreds we do not have such luxury. Moreover, for
bin sizes of interest, even the $m_i$ may not always be large enough
for shot noise to be negligible. The solution is to view {\it both\/}
the sets $\(m_i)$ and $\(s_i)$ as samples drawn from some underlying
$f$ that is constant within bins. The likelihood then takes the form
$$ \cprob({\rm data}|{\rm model}) \propto W = \W.  \putnum$$
   \nameeq{Wprop}
This formula is derived in the Appendix, but note two intuitively
desirable properties of $W$: (i)~the $(m_i+s_i)!$ factor favors large
$m_i$ coinciding with large $s_i$, but the denominator discourages
extremes like $m_i=M$ at the bin with highest $s_i$ and 0 elsewhere;
and (ii)~if some outlier observation lands in a bin with no model
points (i.e., $m_i=0,s_i=1$), that bin contributes unity to the
product---in this sense $W$ is robust against outliers.

Although the formula (\\{Wprop}) is symmetric in $m_i$ and $s_i$,
operationally these two sets of numbers will play quite different
roles. The $s_i$ derive from data and, for a given data set and
binning, they are fixed.  The $m_i$, on the other hand, depend on the
scaling and orientation parameters and will vary as the those
parameters are adjusted to maximize $W$.

To explain the details of the use of $W$ it is probably best to work
through an example, and below we work through the problem of scaling
and orienting \Nbody\ models of the Milky Way bulge and inner disc
from $l,b,v$ measurements.  As it happened, it was this problem that
led to the present work, but the bulge is a good example to illustrate
anyway, for two reasons. Firstly, it is a triaxial system with the
interesting complication that its depth is not negligible compared to
its distance.  Secondly, there are several both of data sets (te
Lintel Hekkert \etal\ 1991, Beaulieu 1996, Sevenster \etal\ 1997a,b)
and models (Sellwood 1993, Zhao 1996, Fux 1997) in the recent
literature.

\section 2. Example: Sellwood's Galactic Model

Figure~1 shows an \Nbody\ model of the Galaxy by Sellwood (1993). Note
the bar in the bulge, which makes the bulge triaxial.  The real
Galactic bulge is now generally agreed to have a substantial bar
(oriented such that the side nearer to us is receding); see Gerhard
(1996) for a review.  Hence the interest of comparing models such as
Sellwood's with kinematic data.

\fig[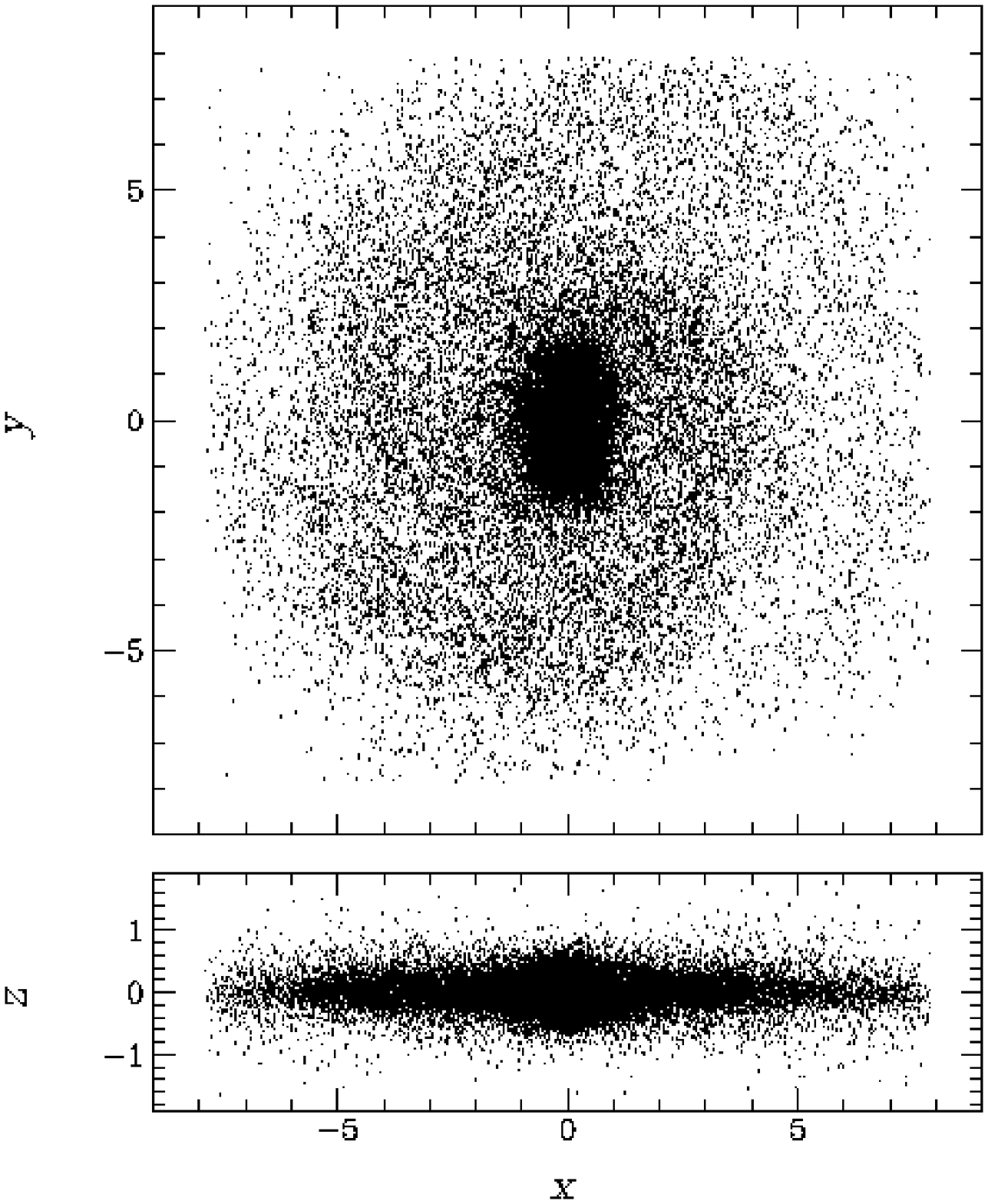,\hsize,\hsize] 1. Spatial positions of 43802 particles in
Sellwood's (1993) model for the Galaxy.  As the spiral arms in the
upper panel suggest, the model rotates counterclockwise, so positive
$z$ corresponds to the South Galactic direction.  The mock $l,b,v$
data are generated by viewing the particles from a 7~o'clock position
on the $(x,y)$ plane at radius of 6.

Let us put ourselves at the point $(-R_0\sin\varphi,R_0\cos\varphi,0)$
in the \Nbody\ model and look towards the bulge.  To evaluate the
observed quantities from this location, we rotate and scale the model
thus:
$$ \eqalign{& x' = x\cos\varphi - y\sin\varphi \cr
            & y' = x\sin\varphi + y\cos\varphi + R_0 \cr
            & z' = z \cr  \noalign{\smallskip}
            & v_x' = \vscl (v_x\cos\varphi - v_y\sin\varphi) - v_0 \cr
            & v_y' = \vscl (v_x\sin\varphi + v_y\cos\varphi) \cr
            & v_z' = \vscl v_z \cr}   \putnum$$  \nameeq{scaleq}
and then compute
$$ \eqalign{& {r'}^2 = {x'^2}+{y'}^2+{z'}^2 \cr
            & l = \arctan(y',x') \cr
            & b = \arcsin(z'/r') \cr
            & v = (x'v_{x'}+y'v_{y'}+z'v_{z'})/r'. \cr}
   \putnum$$ \nameeq{lbveq}
Here $R_0$ is our Galactocentric radius in model units (Sellwood
suggests $R\simeq6$), $\vscl$ is the scaling factor between real and
model velocity units, $v_0$ is our tangential velocity in real units,
$\varphi$ is the viewing angle of the bar, and our radial velocity is
assumed 0 or corrected for.  In the convention implied by equations
(\\{scaleq}) and (\\{lbveq}), $\varphi$ between 0 and $90^\circ$ means
that the nearer side of the bar is at positive $l$ and (because the
model has positive rotation) positive $v$. The real Galactic bar is
believed to be in such an orientation.

The asymmetry between the spatial and velocity parts of equation
(\\{scaleq}) may seem odd---why not
$$ \eqalign{& x' = \rscl(x\cos\varphi - y\sin\varphi) \cr
            & y' = \rscl(x\sin\varphi + y\cos\varphi) + R_0 \cr
            & z' = \rscl z ?}  \putnum$$
We {\it would\/} need an extra parameter $\rscl$ if we were
considering proper motion data (available for Baade's window stars in
Spaenhauer \etal\ 1992).  But from equation (\\{lbveq}) the observables all
depend only on the ratio $R_0/\rscl$, so in equation (\\{scaleq}) we drop
$\rscl$.  Then $R_0$ in effect becomes a surrogate for the spatial
scale: if our true Galactocentric distance is 8.5\thinspace kpc then
$$ \rscl = \left(8.5\over R_0\right) \rm \left(kpc\over model\
unit\right). $$
With proper motion data, in principle $\rscl$ and
$R_0$ could both be determined, thus providing the actual
Galactocentric distance; but with only $l,b,v$ that distance must be
supplied separately to get $\rscl$.  Once we have $\rscl$, we can get
the mass because the scale for $G\times\rm mass$ is
$\rscl\times\vscl^2$ (since $GM$ has dimensions of $L^3T^{-2}$).

Consider now a survey of $l,b,v$ measurements, which we would like to
compare with the simulation and infer $R_0$, $\varphi$, $\vscl$, and
$v_0$ to the extent possible.  The first step is to choose the bins in
$l,b,v$ for comparison---more on choosing bins below, but for the
moment suppose we have chosen our $B$ bins.  This sets the $\(s_i)$.
The $\(m_i)$ will depend on what scaling and orientation parameters we
choose; for any choice we can put the model particles through the
transformations (\\{scaleq}) and (\\{lbveq}), bin them up, and
randomly pick $M$ out of all the particles that fall into our $B$
bins, thus getting the $\(m_i)$.  Then we calculate $W$, which clearly
depends on the parameters.\note{The number of model particles that
fall into our $B$ bins will depend on the parameters. Particles may
fall outside the survey region where we might have no bins.  But $M$
must be kept the same for all parameter values, i.e., we must always
choose a subset of size $M$ of those model particles that {\it do\/}
fall into our $B$ bins.  Otherwise the formula (\\{Wprop}) for $W$
becomes invalid (see the Appendix).}  Clearly, our strategy to
estimate the parameters will be to vary them so as to maximize $W$.
Getting error bars and testing the model are a little more involved,
and discussed in detail in Section~3.  They will involve simulating
data sets from the model.  Generating a simulated data set $\(s_i)$
from the model is like generating $\(m_i)$, except that we choose $S$
particles rather than $M$.  Sometimes we will be calculating $W$ for
two sets of occupancies $\(s_i)$ and $\(m_i)$, both of which come from
the model, but using different parameter values.\note{If we are going
to compare mock data generated from a model with that model, it is
important to then remove from the model those points which went into
the mock data.  A model particle should never contribute to both
$\(s_i)$ and $\(m_j)$ at the same time.  The reason is that the data
are not supposed to correspond {\it exactly\/} to any model particles,
only to have come from the same distribution function.}

How to choose the bins?  Because of the assumptions that go into the
derivation of $W$, it is best to avoid unequally sized bins.  But
there is no need to have bins in unsurveyed regions, so bins need not
be contiguous. The bin size requires some thought.  I cannot suggest
any definite prescription for the bin size, but there are two
guidelines. Firstly, $B$ should be several times smaller than $M$, so
that the $m_i$ can be large enough to actually carry some information
about the distribution function; $B\simeq M/5$ seems serviceable.
There is no problem with $B\gg S$; after all, the continuous limit is
$M\gg B\gg S$.  Secondly, the binning should not be so coarse that is
misses important features in the distribution function. Too coarse a
binning can lead to strange biases, as the following suggests. The
scale height of the bulge (as seen from the solar system) is about
$2.2^\circ$; suppose the bins were $5^\circ$ in $b$.  When fed data
binned thus, any model fitting procedure is likely to respond by
fitting a model with an increased scale in $b$.  An easy way to do
this is to increase $R_0$---but then the fit would have to compensate
for the scale in $l$, which it might do by reducing $\varphi$ to make
the bar more nearly end-on; but a nearly end-on bar will tend to give
larger $v$ values, and this in turn might be compensated by reducing
$\vscl$.

\section 3. Use of the {\eighti W} function

It is straightforward to incorporate the likelihood $W$ in standard
Monte-Carlo procedures for parameter estimation and model testing, and
the following describes how this can be done.  The approach here is
not the only possible one, and Bayesian purists would reject it
entirely; but it seems computationally the most tractable.

It is helpful to consider two functions, $D$ and $\Omega$.

$D(\omega)$ is a function that generates a data set from the model
using parameter values $\omega$.  $D$ is probabilistic, so there can
be many possible data sets $D_1,D_2,\ldots(\omega)$ from the same
model and parameters.  If the model is correct and the true
values are $\omt$, then the observed data can be thought of as one
realization of $D$:
$$ \Dobs = D(\omt).  \putnum$$

$W$ depends on both the data and the parameters:
$$ W = W(D,\omega') \qquad{\rm or:}\ W(D(\omega),\omega').
\putnum$$
In general $\omega'\neq\omega$; $\omega$ leads to the data and hence
to the $\(s_i)$, but the $\(m_i)$ are got by applying a possibly
different value $\omega'$ to the model.  We now define the function
$\Omega$ thus:
$$ \Omega(D) = \hbox{$\omega'$: $W(D,\omega')$ is maximum}.  \putnum$$
To calculate $\Omega(D)$, we don't need to know what $\omega$ value
gave $D$; but $\Omega(D)$ is an estimator for that unknown value
(in fact, a maximum likelihood estimator, because $W$ is a
likelihood).\note{If we had
$$ \big\langle\Omega(D(\omega))\big\rangle = \omega, $$
(the average being over an ensemble of $D_n$ with $\omega$ held fixed)
then $\Omega$ would be an unbiased estimator.  As suggested in
Section~1, in practice $\Omega$ will have some bias, because for one
thing the binning process introduces bias. But that is not a problem
provided we can test for bias and correct for it where required.}

To estimate parameters we calculate $\omest=\Omega(\Dobs)$.  For error
bars on $\omest$ we want the scatter in $\Omega(D(\omt))$.  But since
in real applications we won't know $\omt$, we can take instead the
scatter in $\Omega(D(\omest))$; from this we can read off desired
confidence limits.  This is standard Monte-Carlo error
estimation---see Figures 15.6.1 and 15.6.2 in {\sl Numerical
Recipes\/} by Press \etal\ (1992).

Testing the model is a little more complicated.  We need some
statistic that measures the goodness of a parameter fit, but the well
known ones do not help us: $\chi^2$ is inappropriate for the reasons
given in Section~1, and KS and its relatives are inapplicable because
the data are not one-dimensional.  However, there is an obvious choice
of statistic: $W$ itself.  To apply it, we compare $W(\Dobs,\omest)$
with the distribution of $W(D(\omt),\omest)$.  If $W(\Dobs,\omest)$
lies in the lowest percentile of the distribution of
$W(D(\omt),\omest)$, then the model is rejected at 99\% significance,
and so on.  For $\chi^2$ and also for KS and its relatives, the
distribution corresponding to $W(D(\omt),\omest)$ is model
independent.  In our case we will need to calculate the distribution;
but again, we won't know $\omest$, so we will have to substitute the
distribution of $W(D(\omest),\omest)$.

Note that if the goodness of fit test leads to rejection of the model,
parameter estimates from that model must be rejected too (even if the
error bars claim high accuracy).

The main algorithmic problem is locating the maximum of $W$ in the
multi-dimensional parameter space of $\omega$.  My implementation
basically follows Charbonneau (1995).  Programs are available to
anyone interested.

\section 4. Simulations with Sellwood's model

In this section I present some Monte-Carlo simulations to gain some
idea of which bulge parameters can be constrained from current surveys
and how well.

\fig[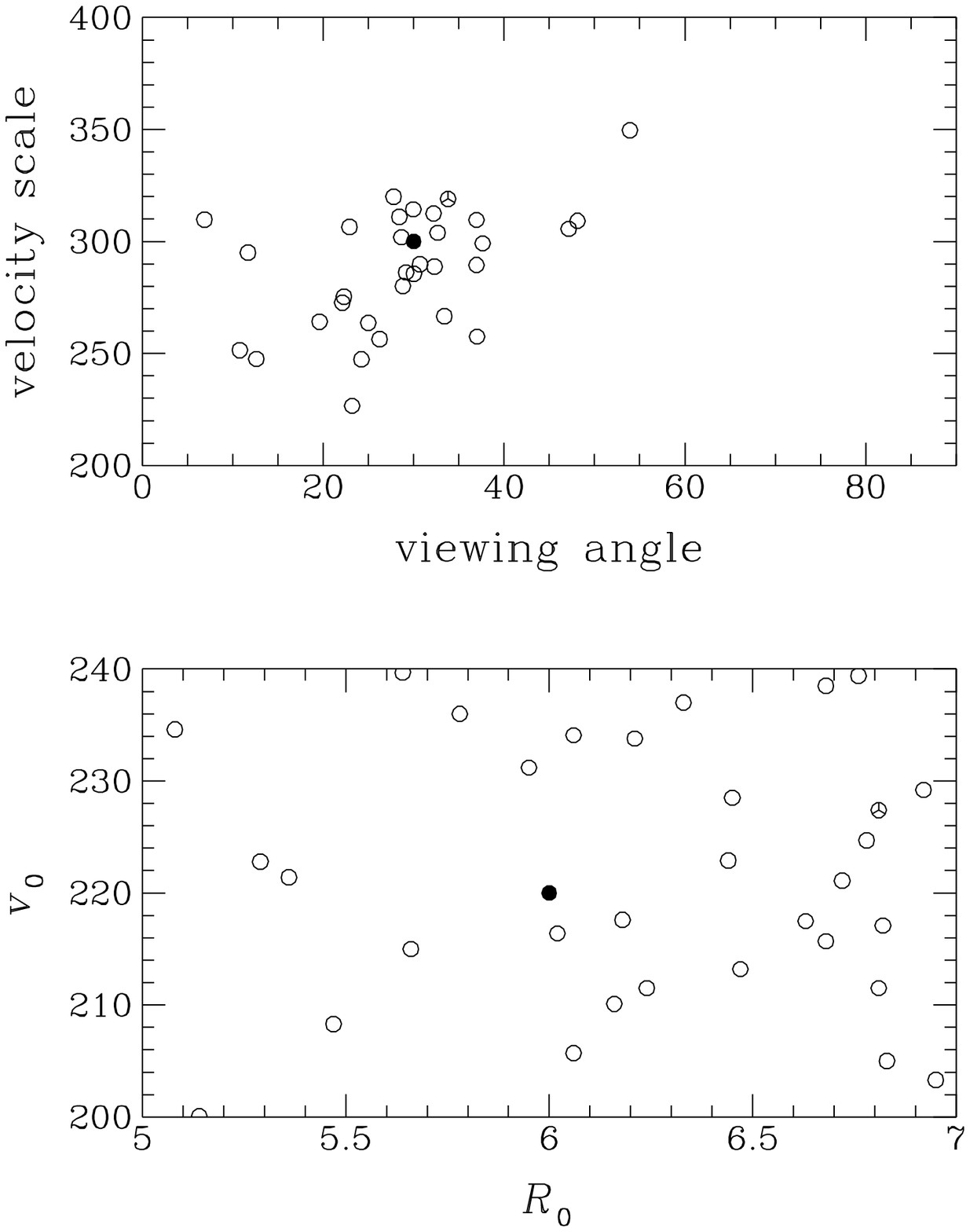,\hsize,\hsize] 2. Recovery of parameter values from
mock surveys of 300 objects in the range $|l|<10.5^\circ$,
$|b|<3.25^\circ$.  The filled circles show the parameter values
used to generate the mock data, i.e., $\omt$.  The open circles show
the parameters recovered from 32 different mock surveys, i.e.,
$\Omega(D_0.\,.D_{31}(\omt))$. The open circles with three cuts inside
refer to the mock survey $D_0(\omt)$ that happened to be first; this
$D_0$ is used again for Figures~3 and 4.

Consider Sellwood's model with $\omt$ being $R_0=6\rm\,model\ units$,
$\varphi=30^\circ$, $v_0=220\rm\,km/sec$, and
$\vscl=300\rm\,km/sec/model\ unit$.  All these are plausible values,
and using them I computed $\Omega(D(\omt))$ for 32 mock surveys, each
having 300 objects in the range $|l|<10.5^\circ$, $|b|<3.25^\circ$,
$|v|<320\rm\,km/sec$.  The size and extent mimics the symmetric part
of the survey of bulge OH/IR stars by Sevenster \etal\ (1997a,b).
(The survey region need not be symmetric---see below.)  I used
$21\times13\times16$ bins in $l,b,v$.  Figure~2 shows the results: the
ranges of the axes in $R_0$, $v_0$, and $\varphi$ are the ranges I
searched; for $\vscl$ I searched the range 0--500.

\fig[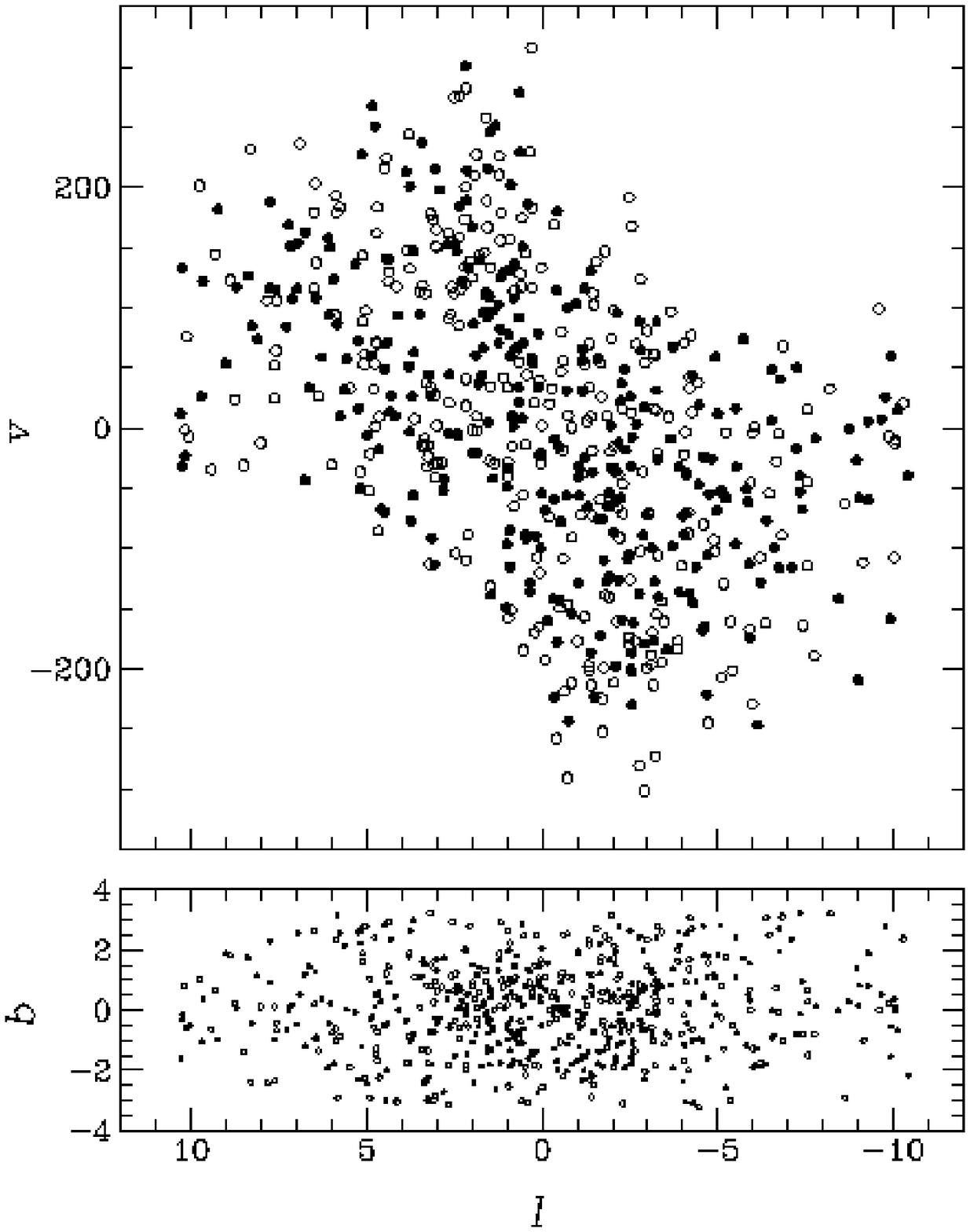,\hsize,\hsize] 3. The open circles show $l,b$ and $l,v$
for a mock survey $D_0(\omt)$. The filled circles show $l,b$ and $l,v$
for a mock survey $D_1(\omest)$, where $\omest=\Omega(D_0(\omt))$.
With a real survey one would use $\Dobs$ instead of $D_0(\omt)$.

Figure 3 shows a sort of direct visual data-model comparison for one
(the first) of the mock surveys.  Such plots are useful in checking
for programming goofs, but otherwise there is not much information one
can extract from them.

Figure 4 illustrates that the maximum of $W$ of the first of the mock
surveys is typical of the values should expect from this model, as
expected since the mock survey came from the model.  With real data
such a plot would test whether the data could plausibly have come from
the model being studied.

From the results in Figure~2, the medians and 68\% range
in $\Omega(D(\omt))$ are $\varphi=(29^{+8}_{-7})^\circ$,
$\vscl=290^{+20}_{-33}$.  These numbers indicate the sort of bias and
error bars we can expect from a survey of this size and extent. We see
that $\vscl$ can be estimated to $\sim10\%$ and $\varphi$ to
$\sim10^\circ$ with no need to correct for bias.  On the other hand,
we get no useful information on $v_0$ and $R_0$.  That $v_0$ is not
constrained by such data is not surprising, since it almost
perpendicular to what is measured.  But the inability to infer $R_0$
is puzzling, especially considering the impressively tight constraint
on $\varphi$.  Evidently, we must use integrated light to estimate
$R_0$.  

It is interesting to see what happens as the surveys get bigger.
Figure 5 shows $\Omega(D(\omt))$ for mock surveys when the size is
extended to 500 and the $l$ range to $-45^\circ<l<10.5^\circ$, which
mimics the full size and extent of Sevenster \etal's (1997a,b) OH/IR
survey.  We now find approximate medians and 68\% ranges of
$\varphi=(29^{+8}_{-8})^\circ$ and $\vscl=290^{+23}_{-11}$, but the
outliers are noticeably less distant.  We also being to notice a bias
towards low estimates for $\vscl$ and high estimates for $R_0$; if the
survey size is increased further, the scatter in $\Omega(D(\omt))$
reduces further, and the biases in $\vscl$ and $R_0$ become
correspondingly more noticeable.

\fig[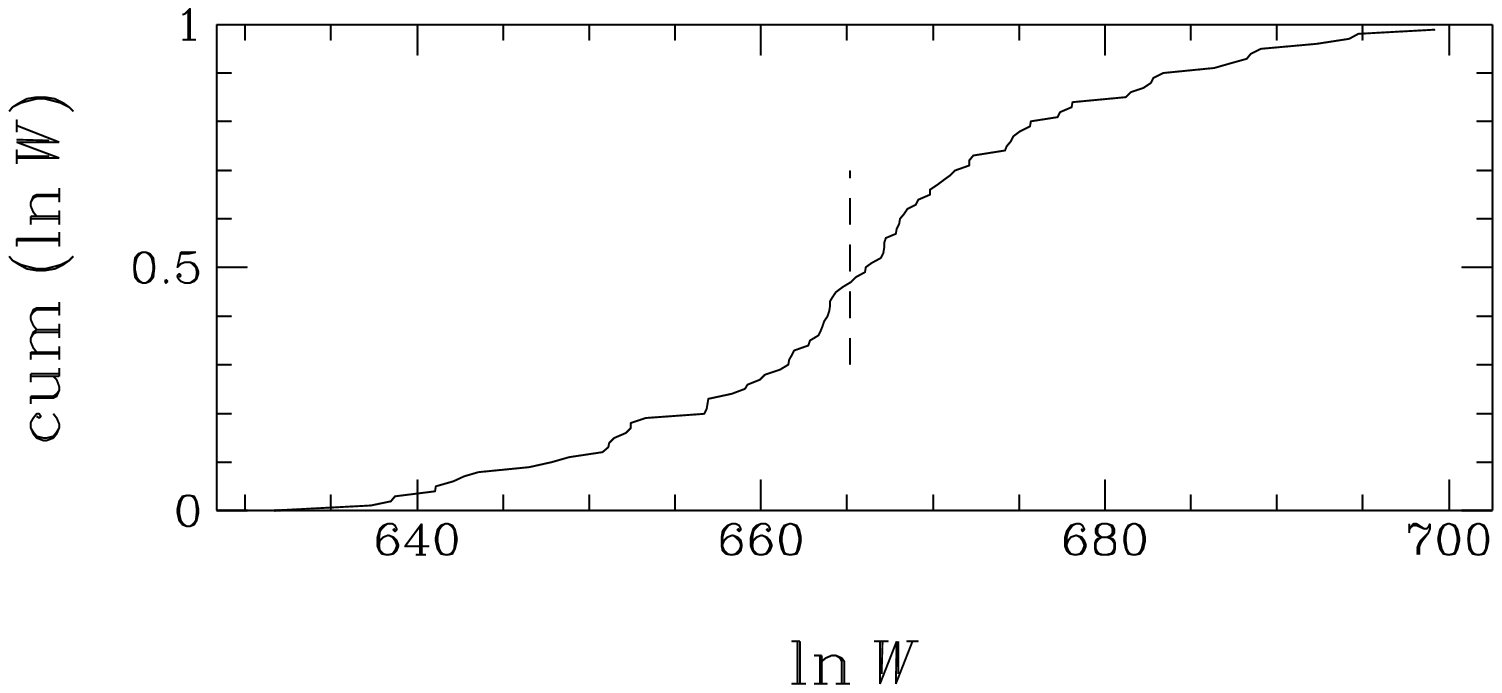,\hsize,.4\hsize] 4. The vertical dashed
line in this figure shows the value of $\ln W(D_0(\omt),\omest)$ where
$\omest=\Omega(D_0(\omt))$ which is marked in Figure~2. With a real
survey $D_0(\omt)$ would be replaced by $\Dobs$. The rising curve is
the cumulative distribution of $\ln W(D_1.\,.D_{100}(\omt),\omest)$.
As expected, $D_0(\omt)$ appears as typical of the distribution
$D_n(\omt)$.

\fig[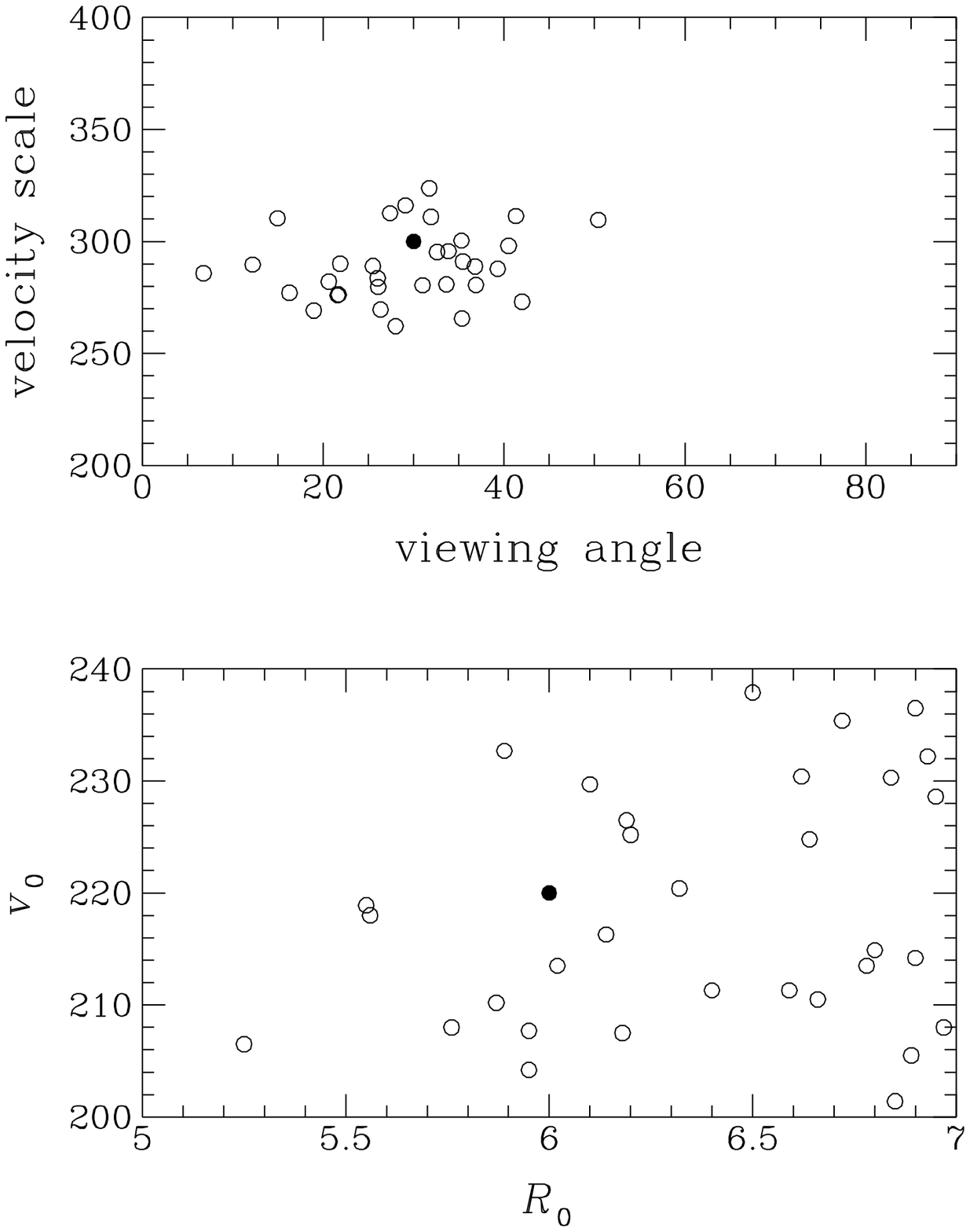,\hsize,\hsize] 5. Similar to Figure~2, but this time
the mock surveys have 500 objects each in the range
$-45^\circ<l<10.5^\circ$, $|b|<3.25^\circ$.

To summarize the results, these survey simulations indicate that
current surveys can constrain the viewing angle of bulge simulations
to $<10^\circ$ and the velocity scale to $<10\%$ (at 68\% confidence).
The spatial scales will need to be set independently using integrated
light.  Kalnajs (personal communication) obtains $\sim10\%$ or better
constraints on $R_0$ by comparing \Nbody\ models and integrated light
from COBE.  Combining with $<10\%$ uncertainties on $\vscl$, it
appears that \Nbody\ models could be scaled in mass to $\sim25\%$. The
resulting predictions for microlensing optical depths would easily be
tight enough for interesting confrontations with bulge microlensing
observations.

\bigskip\noindent I am grateful to Ken Freeman for posing this
problem and to Sylvie Beaulieu and Maartje Sevenster for teaching me
about bulge observations.  Thanks also to Agris Kalnajs and Maartje
Sevenster for the appropriate mixture of enthusiasm and skepticism
about $W$.

\vfill\supereject

\begingroup
\input refs

\endgroup

\eqnum=0\def\theeqnum{\rm A\the\eqnum}
\def\nameeq#1{\sdef{#1}{\theeqnum}\ignorespaces}
\section Appendix 

This appendix derives the likelihood formula (\\{Wprop}) from
probability theory arguments.

Continuing the notation of Section~1, we suppose that there are
$B$ bins in all, and that in the $i$-th bin $f=f_i$, and
both $\(m_i)$ and $\(s_i)$ are drawn from $\(f_i)$.

The joint probability for the bin occupancies is the product of two
multinomial distributions:
$$ \cprob(s_i,m_i|f_i) = M!\,S!\boxes {f_i^{m_i+s_i}\over m_i!s_i!}.
   \putnum$$
Equation (\prevenum1) could be used to estimate the $f_i$, with
uncertainties.  But as we have no particular interest in the $f_i$ as
such, we marginalize them out in the usual way, which is to integrate
over all allowed values of the $f_i$---that means all combinations of
values of the $f_i$ between 0 and 1, subject to $\sum_if_i=1$.
Distributions of $f_i$ that most resemble the $m_i$ and $s_i$ will,
according to (\prevenum1), contribute most to the integral.
Marginalization is just an application of the additive rule for
probabilities.  Using the identity
$$ \left(\boxes\int\! f_i^{n_i}\,df_i\right) \,
   \delta\left({\textstyle\sum_j f_j-1}\right)
   = {1\over (N+B-1)!} \boxes n_i! \putnum$$
we get
$$ \prob({\(s_i)},{\(m_i)}) = {M!\,S!\,(B-1)!\over (M+S+B-1)!}\W.
   \putnum$$
Putting $S=0$ in (\prevenum1) gives $\prob({\(m_i)})$, and hence
$$ \cprob({\(s_i)}|{\(m_i)}) = {S!\,(M+B-1)!\over (M+S+B-1)!} \W.
   \putnum$$ \nameeq{fullW}
If $B=1$ then the probabilities in (\prevenum2) and (\prevenum1)
become unity, as they should. For the sort of applications of interest
in this paper $M$, $S$, and $B$ are fixed, so we can ignore the
normalization in (\\{fullW}) and work only about $W$ as defined in
equation (\\{Wprop}).

If $M$ is large enough compared to $S$ that $m_i+s_i\simeq m_i$ and
(\\{fullW}) simplifies to
$$ \cprob({\(s_i)}|{\(m_i)}) \propto \prod_{i} {m_i^{s_i}\over s_i!},
   \putnum$$
which amounts to saying that $f_i=m_i/M$ because the shot noise in the
$m_i$ is negligible.  If $M$ is large enough that we can make the bins
so small that each $s_i=0$ or 1, but still all $m_i\gg1$, then
(\prevenum1) simplifies further, to
$$ \cprob({\(s_i)}|{\(m_i)}) \propto \prod_{j:s_j=1} m_j,   \putnum$$
which amounts to the formula (\\{likelicont}) for the continuous case.
Thus, (\\{fullW}) has the expected large-$m_i$ limits.

\closeout1 \vfill\eject
               \leftline{\bf\uppercase{figure captions}}
               \openout1=pscalls.tmp \input captions.tmp \closeout1
               \vfill\supereject \input pscalls.tmp \end

%% file: macros.tex
\def\ifundef#1{\expandafter\ifx\csname#1\endcsname\relax}


\newif\ifnoopshun\noopshuntrue

\ifundef{draft}\else\message{Draft style}\noopshunfalse\fi
\ifundef{preprint}\else\message{Preprint style}\noopshunfalse\fi
\ifundef{submission}\else\message{Submission style}\noopshunfalse\fi

\ifnoopshun
 \def\prep{prep } \def\press{press }  \def\prel{draft }
 \immediate\write0{}
 \message{This can be TeXed in three different styles:}
 \message{"prep"  preprint style with single-spaced text;}
 \message{"press" style---double-spaced with figures at the end;}
 \message{"draft" style---like prep but in small two-page landscape format;}
 \immediate\write0{}
 \message{Please type prep, press or draft: }
 \loop
    \read-1 to \opshun
    \ifx\opshun\press\noopshunfalse
        \message{OK, press style it is...} \fi
    \ifx\opshun\prep\noopshunfalse
        \message{OK, preprint style it is...} \fi
    \ifx\opshun\prel\noopshunfalse
        \message{OK, draft style it is...} \fi
    \ifnoopshun
        \message{I don't understand "\opshun"!}
        \message{prep or press or draft please: }
    \repeat
 \immediate\write0{}
\fi

\ifundef{draft}\magnification=1200\fi


\input ./fonts

\input epsf

\ifundef{Afour} \else \vsize=24.65truecm \hsize=15.94truecm \fi
\ifundef{draft} \else \input landscap \fi

\def\psfigcall#1#2{\def\epsfsize##1##2{#2}\centerline{\epsfbox{#1}}}

\def\insfig[#1,#2,#3]#4.{\midinsert\parindent=0pt
           \vbox to #3 {\vss\psfigcall{#1}{#2}}
           \def\par{\endgraf\endinsert} \ninepoint
           \ignorespaces {\bf Figure#4.}\quad \ignorespaces}

\def\blackout{\catcode`\\=12\catcode`\{=12\catcode`\}=12\catcode`\$=12
\catcode`\&=12\catcode`\#=12\catcode`\^=12\catcode`\_=12\catcode`\@=12
\catcode`\~=12\catcode`\%=12\catcode`\|=12}

{\obeylines \global\let
                       \relax
    \gdef\fig{\begingroup\obeylines\verbatim}
    \gdef\verbatim{\blackout\def
                   ##1
                   {\setbox0\hbox{##1}%
                    \ifdim\wd0<1pt \endgroup \else\immediate\write1{##1}\fi
                   }
   \string\figcap}}

\def\figcap[#1,#2,#3]#4.{\write1{\null\vfil
           \vbox to #3{\vss\string\psfigcall{#1}{#2}}}
    \write1{\bigskip\bigskip\centerline{\twelvebf Figure#4}\vfil\string\eject}
    \bigskip\noindent{\bf\ignorespaces#4.}}

\expandafter
\ifx\csname submission\endcsname\relax
      \let\adjustlinespacing\relax
      
      \let\fig=\insfig
\else \def\adjustlinespacing{\baselineskip=1.5\baselineskip}
      \adjustlinespacing
      
      \openout1=captions.tmp
\fi

\def\maybesupereject{\expandafter\ifx\csname submission\endcsname\relax
                     \vfill\supereject\fi}
\catcode`\@=11

\newcount\notenum

\def\vfootnote#1{\insert\footins\bgroup\eightpoint
     \interlinepenalty=\interfootnotelinepenalty
     \splittopskip=\ht\strutbox \splitmaxdepth=\dp\strutbox
     \floatingpenalty=20000
     \leftskip=0pt \rightskip=0pt \parskip=1pt \spaceskip=0pt \xspaceskip=0pt
     \smallskip\textindent{#1}\footstrut\futurelet\next\fo@t}

\def\note{\global\advance\notenum by 1
    \edef\n@tenum{$^{\the\notenum}$}\let\@sf=\empty
    \ifhmode\edef\@sf{\spacefactor=\the\spacefactor}\/\fi
    \n@tenum\@sf\vfootnote{\n@tenum}}


\def\iterate{\let\next=\relax\body\let\next=\iterate\fi\next}

\def\mydef#1#2{\expandafter\ifx\csname#1\endcsname\relax
               \expandafter\def\csname#1\endcsname{#2}\else
               \errmessage{#1 is already \csname#1\endcsname}\fi}
\newif\ifneedrerun
\def\\#1{\expandafter\ifx\csname#1\endcsname\relax
         ??\needreruntrue \else \csname#1\endcsname\fi}

\def\parb{\par }
\openin1=names
\loop
\ifeof1\else
   \read1 to \cs
   \ifx\cs\parb\else \expandafter\mydef\cs \fi
\repeat

\immediate\openout2=names
\def\sdef#1#2{\expandafter\xdef\csname#1\endcsname{#2}%
              \immediate\write2{{#1}{#2}}}


\newcount\eqnum
\def\nextenum{\global\advance\eqnum by 1 \theeqnum}
\def\prevenum#1{{\advance\eqnum by -#1 \advance\eqnum by 1 \theeqnum}}
\def\nameeq#1{\sdef{#1}{\the\eqnum}\ignorespaces}
\def\putnum{\eqno(\nextenum)}
\def\theeqnum{\rm\the\eqnum}


\def\<#1>{{\left\langle#1\right\rangle}}

\def\witchbox#1#2#3{\hbox{$\mathchar"#1#2#3$}}
\def\leqsim{\mathrel{\rlap{\lower3pt\witchbox218}\raise2pt\witchbox13C}}
\def\geqsim{\mathrel{\rlap{\lower3pt\witchbox218}\raise2pt\witchbox13E}}

\let\twiddle=~ \def~{{}\ifmmode\widetilde\else\twiddle\fi}
\let\comma=\, \def\,{\ifmmode\comma\else\thinspace\fi}

\def\frac#1/#2{\mathchoice  {\hbox{$#1\over#2$}} {{#1\over#2}}
                {\scriptstyle{#1\over#2}}
                {#1\mskip-1.5mu/\mskip-1.5mu#2}}

\def\pderiv(#1/#2){\mathchoice{\partial#1\over\partial#2}
    {\partial#1/\partial#2} {\partial#1/\partial#2} {\partial#1/\partial#2}}
\def\pderivp(#1/#2){\left(\pderiv(#1/#2)\right)}

\def\tderiv(#1/#2){\mathchoice{d#1\over d#2}{d#1/d#2}
                   {d#1/d#2}{d#1/d#2}}
\def\tderivp(#1/#2){\left(\tderiv(#1/#2)\right)}

\def\leftmatrix#1{\null\,\vcenter{\normalbaselines\m@th
    \ialign{$##$\hfil&&\quad$##$\hfil\crcr
      \mathstrut\crcr\noalign{\kern-\baselineskip}
      #1\crcr\mathstrut\crcr\noalign{\kern-\baselineskip}}}\,}

\def\rightmatrix#1{\null\,\vcenter{\normalbaselines\m@th
    \ialign{\hfil$##$&&\quad\hfil$##$\crcr
      \mathstrut\crcr\noalign{\kern-\baselineskip}
      #1\crcr\mathstrut\crcr\noalign{\kern-\baselineskip}}}\,}


\headline={\ifnum\pageno=0\hfil
           \else\ifodd\pageno\hfil\tenit\rhead\quad\hbox{\tenbf\folio}%
                \else\hbox{\tenbf\folio}\quad\tenit\lhead\hfil\fi
           \fi}

\footline={\hfil}

\parskip \smallskipamount

\def\maybebreak#1{\vskip 0pt plus #1\vsize \penalty -1000
                  \vskip 0pt plus -#1\vsize}

\outer\def\section#1\par{\maybebreak{.1}\bigskip\bigskip\bigskip
      \centerline{\eightrm\uppercase{#1}}
      \medskip\noindent}

\def\comment{$\{$\bgroup\eightpoint\it\aftergroup\endcomment\let\next}
\def\endcomment{$\}$}

\def\ordsuffix#1{\ifcase#1\or st\or nd\or rd\or th\or th\or 
    th\or th\or th\or th\or th\or th\or th\or th\or th\or 
    th\or th\or th\or th\or th\or th\or st\or nd\or rd\or 
    th\or th\or th\or th\or th\or th\or th\or st\fi}
\def\dayord#1{#1$^{\rm\ordsuffix{#1}}$}
\def\today{\expandafter\dayord{\number\day}
 \ifcase\month\or
 January\or February\or March\or April\or May\or June\or
 July\or August\or September\or October\or November\or December\fi
 \space \number\year}

\catcode`\@=12

\def\pseudocode{$$\vcenter\bgroup\dimen1=\parindent
        \advance\hsize by -2\dimen1
        \rightskip=\dimen1\parindent=0pt
        \def\begin{\par\advance\leftskip by \dimen1}
        \def\nigeb{\par\advance\leftskip by -\dimen1}
        \setbox1\hbox{$\langle\,$}
        \def\<{\hangindent\wd1\hangafter1$\langle\,$}
        \def\>{$\,\rangle$\par}}
\def\endcode{\egroup\eqno\vrule height0pt width.1pt depth0pt
                         \llap{(\nextenum)}$$}


\def\etal{{\it et al.}}

\def\Nbody{$N$\kern-.08ex-body}
\def\vscl{v_{\rm scale}}  \def\rscl{r_{\rm scale}}

\def\prob(#1){\mathop{\rm prob}\left({#1}\right)}
\def\cprob(#1|#2){\mathop{\rm prob}\left({#1}\!\mid\!{#2}\right)}
\def\boxes{\prod_{i=1}^B}
\def\W{\boxes{(m_i+s_i)!\over m_i!s_i!}}
\def\likeli{\cprob({\rm data}|f)}
\def\(#1){#1}

\def\omt{\omega_{\rm true}}  \def\omest{\omega_{\rm est}}
\def\Dobs{D_{\rm obs}}

%% file: fonts.tex
 at 10 truept
\font\fourteenrm=cmr10 scaled 1440
\font\fourteeni=cmmi10 scaled 1440
\font\twelvebf=cmbx10 scaled 1200

\font\ninerm=cmr9
\font\ninei=cmmi9
\skewchar\ninei='177
\font\ninesy=cmsy9
\skewchar\ninesy='60
\font\nineit=cmti9
\font\ninesl=cmsl9
\font\ninebf=cmbx9
\font\ninett=cmtt9
\def\ninepoint{\textfont0=\ninerm \scriptfont0=\sevenrm 
              \def\rm{\fam0\ninerm}\relax
              \textfont1=\ninei \scriptfont1=\seveni 
              \def\mit{\fam1}\def\oldstyle{\fam1\ninei}\relax
              \textfont2=\ninesy \scriptfont2=\sevensy 
              \def\cal{\fam2}\relax
              \textfont3=\tenex \scriptfont3=\tenex 
              \def\it{\fam\itfam\nineit}\relax
              \textfont\itfam=\nineit
              \def\sl{\fam\slfam\ninesl}\relax
              \textfont\slfam=\ninesl
              \def\bf{\fam\bffam\ninebf}\relax
              \textfont\bffam=\ninebf \scriptfont\bffam=\sevenbf
              \def\tt{\fam\ttfam\ninett}\relax
              \textfont\ttfam=\ninett
              \setbox\strutbox=\hbox{\vrule
                   height8pt depth3pt width0pt}\baselineskip=11pt
              \adjustlinespacing
              \rm}

\font\eightrm=cmr8
\font\eighti=cmmi8
\skewchar\eighti='177
\font\eightsy=cmsy8
\skewchar\eightsy='60
\font\eightit=cmti8
\font\eightsl=cmsl8
\font\eightbf=cmbx8
\font\eighttt=cmtt8
\def\eightpoint{\textfont0=\eightrm \scriptfont0=\fiverm 
                \def\rm{\fam0\eightrm}\relax
                \textfont1=\eighti \scriptfont1=\fivei 
                \def\mit{\fam1}\def\oldstyle{\fam1\eighti}\relax
                \textfont2=\eightsy \scriptfont2=\fivesy 
                \def\cal{\fam2}\relax
                \textfont3=\tenex \scriptfont3=\tenex 
                \def\it{\fam\itfam\eightit}\relax
                \textfont\itfam=\eightit
                \def\sl{\fam\slfam\eightsl}\relax
                \textfont\slfam=\eightsl
                \def\bf{\fam\bffam\eightbf}\relax
                \textfont\bffam=\eightbf \scriptfont\bffam=\fivebf
                \def\tt{\fam\ttfam\eighttt}\relax
                \textfont\ttfam=\eighttt
                \setbox\strutbox=\hbox{\vrule
                     height7pt depth2pt width0pt}\baselineskip=9pt
                \adjustlinespacing
                \rm}

%% file: title.tex
\tolerance=500

\def\title#1\par{\pageno=0\null
    \line\bgroup\hfil
    \vbox to 8.3cm
    \bgroup\hsize=9.17cm
    \baselineskip=12pt\parindent=0pt\parskip=0pt \obeylines
    \def\,{\vskip1pt}
    \null\vfil\bgroup\fourteenrm\baselineskip=18pt
    \textfont0=\fourteenrm \textfont1=\fourteeni}
{\obeylines
\gdef\authors#1

{\bigskip\egroup}
\gdef\abstract#1

{\vfil\egroup\hfil\egroup\noindent}}

\title

A method for comparing discrete
kinematic data and \Nbody\ simulations

\authors

Prasenjit Saha$^1$ \,
Mount Stromlo and Siding Spring Observatories,
Australian National University,
Canberra ACT 0200, Australia

\abstract

This paper describes a method for quantitatively comparing an \Nbody\
model with a sample of discrete kinematic data.  The comparison has
two stages: (i)~finding the optimum scaling and orientation of the
model relative to the data; and (ii)~calculating a goodness of fit,
and hence assessing the plausibility of the model in vew of the data.

The method derives from considering the data and model both as samples
from some underlying binned distribution function, and applying
probability theory arguments.

As an example, I consider a published \Nbody\ model for the Galactic
bulge and disc, and fictitious $l,b,v$ measurements, and recover (with
error estimates) the spatial and velocity scales of the model and the
orientation of the bar.  The fictitious data are actually derived from
the model by assuming the mass scale and the solar position, but their
size and extent mimics a recent survey of OH/IR stars.  The results
indicate that mass of the bulge and our viewing angle of the bar are
usefully estimable from current surveys.

\vfootnote{$^1$}{Present address: Department of Physics
(Astrophysics), Keble Road, Oxford OX1~3RH, United Kingdom. Email:
{\tt saha@physics.ox.ac.uk}}

\vfil\eject

\def\lhead{Comparing discrete kinematics and simulations}
\let\rhead=\lhead

%% file: refs.tex
\parindent=0pt \frenchspacing
\everypar{\hangindent 20pt \hangafter 1}
 
\section References



Arnaboldi, M., Freeman, K.C., Hui, X., Capaccioli, M., \& Ford, H. (1994) 
{\sl ESO Messenger,} {\bf 76}, 40

Beaulieu, S.F. (1996) Ph.D. thesis, Australian National University

Charbonneau, P. (1995) {\sl Astrophys. J. Supp.,} {\bf 101}, 309

Ciardullo, R., Jacoby, G.H., Dejonghe, H.B. (1993) {\sl
Astrophys. J.,} {\bf 414}, 454

Colless M. \& Dunn A.M. (1996) {\sl Astrophys. J.,} {\bf 458}, 435

Fux, R. (1997) preprint, {\tt astro-ph/9706242}

Gerhard, O.E. (1996) in {\sl Proc. AIU Symp. 169, Unsolved problems of
the Milky Way,} ed. L. Blitz \& P. Teuben (Reidel, Dordrecht)

Hui, X. (1993) {\sl Pub. Astron. Soc. Pacific.,} {\bf 105}, 1011

te Lintel Hekkert, P., Caswell, J.L., Habing, H.J., Haynes, R.F., \&
Norris, R.P. (1991) {\sl Astron. Astrophys. Supp.} {\bf 90}, 327

Merritt, D. (1993) {\sl Astrophys. J.,} {\bf 413}, 79

Merritt, D. (1996) {\sl Astron. J.,} {\bf 112}, 1085

Merritt, D. \& Gebhardt, K. (1994) in {\sl Clusters of Galaxies,
        Proceedings of the XXIXth Rencontre de Moriond,} ed. F. Durret,
        A. Mazure \& J. Tr\^an Thanh V\^an (Editions Frontiere: Singapore)


Merritt, D. \& Tremblay, B. (1993) {\sl Astron. J.,} {\bf 106}, 2229

Meylan, G. \& Mayor, M. (1986) {\sl Astron. Astrophys.,} {\bf 166},
122

Press, W.H., Teukolsky, S.A., Vetterling, W.T., \& Flannery,
B.P. (1992) {\sl Numerical Recipes,} second ed. (CUP)


Sellwood, J.A. (1993) in {\sl Back to the Galaxy,} eds. Holt, S.S., \&
Verter, F. (AIP press)

Sevenster, M.N., Chapman, J.M., Habing, H.J., Killeen, N.E.B., \&
Lindqvist, M. (1997a) {\sl Astron. Astrophys. Supp,} {\bf 122}, 79

Sevenster, M.N., Chapman, J.M., Habing, H.J., Killeen, N.E.B., \&
Lindqvist, M. (1997b) {\sl Astron. Astrophys. Supp,} in press

Sivia, D.S. (1996) {\sl Data Analysis, A Bayesian Tutorial,} (OUP)

Spaenhauer, A., Jones, B.F., \& Whitford, A.E. (1992) {\sl
Astron. J.,} {\bf 103}, 297

Tremblay, B., Merritt, D., \& Williams, T.B. (1995), {\sl
Astrophys. J.,} {\bf 443}, L5

Zhao H.S. (1996) {\sl Mon. Not. Roy. astr. Soc.,} {\bf 283}, 149